\documentclass[11pt]{article}

\usepackage{graphicx}
\usepackage{epstopdf}
\usepackage{amsmath}
\usepackage{amssymb}
\usepackage{amsfonts}
\usepackage{amsthm}
\usepackage[usenames]{color}
\usepackage{array}

\usepackage[
      colorlinks=true,
      linkcolor=blue,
      urlcolor=blue,
      filecolor=blue,
      citecolor=red,
      pdfstartview=FitV,
      pdftitle={},
      pdfauthor={},
      pdfsubject={},
      pdfkeywords={},
      pdfpagemode=None,
      bookmarksopen=true
]{hyperref}

\usepackage{epsfig}

\usepackage{hyperref}

\newcommand{\bea}{\begin{eqnarray}}
\newcommand{\eea}{\end{eqnarray}}
\newcommand{\ba}{\begin{array}}
\newcommand{\ea}{\end{array}}
\newcommand{\ee}{\end{equation}}

\numberwithin{equation}{section}

\begin{document}

\begin{flushright}
\texttt{\today}
\end{flushright}

\begin{centering}

\vspace{2cm}

\textbf{\Large{
Stress tensor correlators of CCFT$_2$ \\ using flat-space holography  }}

  \vspace{0.8cm}

  {\large Mohammad Asadi, Omid Baghchesaraei,  Reza Fareghbal }

  \vspace{0.5cm}

\begin{minipage}{.9\textwidth}\small
\begin{center}

{\it  Department of Physics, 
Shahid Beheshti University, 
G.C., Evin, Tehran 19839, Iran.  }\\

  \vspace{0.5cm}
{\tt m$\_$asadi@sbu.ac.ir ,omidbaghchesaraei@gmail.com, r$\_$fareghbal@sbu.ac.ir}
\\ $ \, $ \\

\end{center}
\end{minipage}


\begin{abstract}
 We use the correspondence between three-dimensional asymptotically flat spacetimes and two-dimensional contracted conformal field theories (CCFTs) to derive the stress tensor correlators of CCFT$_2$. On the gravity side we use the metric formulation instead of the Chern-Simons formulation of  three-dimensional gravity. This method can also be used for  four-dimensional case where there is no Chern-Simons formulation for the bulk theory.  
\end{abstract}

\end{centering}

\newpage



\section{Introduction}

 Extending gauge/gravity duality beyond the AdS/CFT correspondence requires that one proposes appropriate dual field theory for the spacetimes which are not asymptotically AdS. One of the  candidates is  asymptotically flat spacetimes. These spacetimes are given by vanishing cosmological constant limit of the asymptotically AdS counterparts.  This connection on the gravity side may be a hint  to the proposal of  a dual field theory for the asymptotically flat spacetimes. One of the proposals which links the flat-space limit on the bulk side to the ultra-relativistic limit of the boundary theory, was put forward in \cite{Bagchi:2010zz}-\cite{Bagchi:2012cy}. This proposal which we henceforth call  flat/CCFT, suggests a holographic connection between the asymptotically flat spacetimes in $(d+1)$-dimensions and contracted conformal filed theories (CCFT)  in $d$-dimensions. 
 
A CCFT is given by taking an ultra-relativistic limit of the corresponding CFT. In the ultra-relativistic limit the speed of light approaches  zero and  in this singular limit, the symmetries of the theory are not  Poincare symmetry. In two dimensions, the contracted conformal algebra is given  by     Inonu-Wigner contraction of  two copies of the Virasoro algebra. Starting with a CFT$_2$, the contracted algebra is obtained by using  the generators of the Virasoro algebra and then contracting the time-coordinate \cite{Bagchi:2012cy}. The ultra-relativistic limit of the conformal algebra is the opposite of the non-relativistic limit which gives rise to the  Galilean conformal algebra (GCA)\cite{Bagchi:2009my}. In two dimensions, these two algebras are isomorphic but in higher dimensions they are different.  
 
 A symmetry similar to the contracted conformal symmetry also appears as the asymptotic symmetry of the asymptotically                                      flat spacetimes\cite{BMS}-\cite{aspects}. This symmetry which is called the BMS symmetry is infinite\,- dimensional for three and four dimensions.   Taking the flat-space limit of the generators of the AdS asymptotic symmetry leads to the generators of the BMS algebra\cite{Barnich:2012aw}. Thus it is plausible to propose that the ultra\,- relativistic limit of the CFT is indeed, the dual of the flat-space limit in the asymptotically AdS  spacetimes. This idea is used in \cite{Bagchi:2010zz}-\cite{Bagchi:2012cy} where  a holographic duality between the asymptotically flat spacetimes and CCFTs is proposed. 
 
 Holographic calculation of the stress tensor correlators is a good check  for the correctness of the correspondence between a field theory and a gravitational dual  theory. It is well known that the correlation functions of the operators in CFTs have universal forms. One of the successes of the AdS/CFT correspondence is its proposed method for the   holographic calculation of these correlators.
 
 Similar to the AdS/CFT correspondence, the correlation functions of the operators in a CCFT must have a dual description in the asymptotically flat spacetimes. There are two plausible  ways to establish a dictionary which relates  calculations in the two sides of the duality.   One can  ignore the AdS/CFT correspondence and consider flat/CCFT in its own right or one can  take the appropriate limit of the  calculations of the AdS/CFT correspondence. Both of these methods have been invoked and  the results have been consistent, so far.  
 
 Calculating the stress tensor of  CCFT by using  flat-space holography was carried out for the first time  in  \cite{Fareghbal:2013ifa}. The method  used to find the stress tensor of CCFT$_2$, is taking the appropriate limit of the AdS/CFT computations. On the other hand, in  \cite{Bagchi:2015wna} a direct method is invoked  which yields  the correlators of CCFT$_2$. However, the  holographic calculations of correlation functions in the gravity side just performed by using the Chern-Simons formulation of three\,- dimensional flat-space gravity.  Generalizing such a correspondence to the        higher\,- dimensional cases for which there is no Chern-Simons formulation for the gravity theory, necessitates  the metric formulation of such a calculations. 
 
 In the present  paper we use the metric formulation of three\,- dimensional gravity in order to calculate the stress tensor correlators via holography. The fact that stress tensor of a field theory can be used to find the conserved charges of the symmetries is employed to derive an expression for  the stress tensor components in terms of the conserved charges. Then  the flat/CCFT proposal is used and the charges are  substituted by results in the literature, found directly in the flat spacetimes. Our results in this paper are consistent with \cite{Fareghbal:2013ifa}. This method has also been used previously  for the quasi-local stress tensor of the Kerr black hole   \cite{Fareghbal:2016hqr} and the results are consistent with the ones obtained through  taking the flat-space limit.
 
 To  calculate the higher-point correlation functions, we make use of invariance of  the correlators under the action of the global part of  BMS$_3$ algebra. We track this invariance back  to the gravity side and find a general expression for all of the non-zero stress tensor correlator. Our results also confirm the idea that the symmetry algebra  of CCFT$_2$ is so rich that it dictates  a universal form for the correlators. Another non-trivial point in our calculation is assuming a non-symmetric stress tensor for the CCFTs. Our investigations in the gravity side show that  a covariant conservation formula requires a non-symmetric stress tensor. The fact that CCFTs do not exhibit Poincare' symmetry helps us avoid any inconsistencies. Our calculations in the present paper provide yet  another confirmation for the fact that  asymptotically flat spacetimes do have  holographic duals which are CCFTs living in one less dimension.
 
 In Sec.$ 2 $  we introduce the stress tensor of CCFT$_2$ by using holographic method and metric formulation of  three\,- dimensional gravity. In Sec.$ 3 $ we calculate the p-point functions of the stress tensor by using holography. The last section, Sec.$ 4 $,  is devoted to a discussion and to directions for possible  future investigations.

\section{Stress tensor of CCFT}
Our goal is to calculate the correlation functions of the  CCFT$_2$ stress tensor. In the first step we need to introduce the stress tensor. According to our convention,  a CCFT$_2$ is a theory which is defined by the following infinite\,- dimensional symmetry:
\begin{eqnarray}\label{algebra}
 [L_m,L_n]&=&(m-n)L_{n+m}+C_L m (m^2-1) \delta_{m+n,0},\cr  [L_m,M_n]&=&(m-n)M_{n+m}+C_M m (m^2-1) \delta_{m+n,0},
\end{eqnarray}
where $n$ and $m$ can take any integer values. Similar to CFT$_2$, one may expect that  the above  infinite\,- dimensional symmetry  yields some universal results which are independent of the underlying action. The algebra \eqref{algebra} is given by the Inonu-Wigner contraction of the Virasoro algebra. Thus, one may consider CCFT$_2$ as a contracted theory obtained from a parent CFT. There are two possible contractions of the Virasoro algebra which lead to \eqref{algebra}, a non-relativistic and an  ultra\,- relativistic contraction. The first one which is given by taking very large limit of  the light speed, corresponds to scaling $x\to\epsilon x$ and $\epsilon\to 0$. On the other hand the ultra\,- relativistic contraction is obtained by the limit of vanishing light speed or equivalently scaling $t\to\epsilon t$ and $\epsilon\to 0$. In two dimensions both  the non-relativistic and               ultra\,- relativistic contractions of the Virasoro algebra give rise to the same algebra as in  \eqref{algebra}. However,  in general , by CCFT we mean a theory for which the symmetry is given by the ultra-relativistic limit. The non-relativistic limit yields Galilean conformal algebra (GCA) which is interesting on its own\cite{Bagchi:2009my}. 

We suppose that CCFT$_2$ lives on a cylinder with metric
\begin{equation}\label{metric0}
ds^2= -du^2+R^2 d\phi^2
\end{equation}
where $R$ is the radius of the cylinder,  which will be fixed later when we use the holographic dictionary. Our starting point for finding the stress tensor of CCFT is  the formula which gives the conserved charges of symmetry generators $\xi$. Using \eqref{metric0} we can write
\begin{equation}\label{2dcharge}
Q_{\xi}=R\int_{0}^{2\pi} d\phi J^u=R\int_{0}^{2\pi} d\phi T^{u\mu}\xi_\mu, 
\end{equation}
where $J^{\mu}$ is the symmetry current and $T^{\mu\nu}$ is the stress tensor. Here, we do not impose any conditions  on the components of the stress tensor. 
For a CCFT that lives on the cylinder  one can introduce a representation for the generators of \eqref{algebra} 
\begin{equation}
L_n=i e^{i n \phi} \left(\partial_\phi+i n u\partial_u\right),\qquad M_n=i e^{i n \phi}\partial_u.
\end{equation}
Thus we can write 
\begin{align}\label{T in terms of Q}
\nonumber Q_{M_n}&=-iR\int_0^{2\pi} d\phi\, e^{in\phi}\, T^{uu},\\
Q_{L_n}&=R\int_0^{2\pi} d\phi\, e^{in\phi}\, \left(nuT^{uu}+iR^2 T^{u\phi}\right).
\end{align}
Using the orthogonality condition of Fourier modes,
we can find $T^{uu}$ and $T^{u\phi}$ from \eqref{T in terms of Q} as
\begin{align}\label{stress tensor component field theory}
\nonumber T^{uu}&={i\over 2\pi R}\sum_n Q_{M_n} e^{-in\phi}\\
T^{u\phi}&={-i\over 2\pi R^3}\sum_n e^{-in\phi}\left(Q_{L_n}-iun Q_{M_n}\right)
\end{align}
The other components must be determined by using the conservation and traceless-ness conditions. However, in order to check the above calculations and find other components we make use of the flat/CCFT proposal and  first do a holographic calculation.

 \subsection{Holographic calculation using Flat/CCFT correspondence }
 The calculations in the previous section are pure field theoretic ones and we merely defined a two\,- dimensional field theory by its symmetries. However, as is proposed in \cite{Bagchi:2010zz}-\cite{Bagchi:2012cy}  this two\,- dimensional field theory has a holographic  dual theory. The dual theory is three\,- dimensional gravity in asymptotically flat backgrounds. The asymptotic symmetries of such a spacetimes  at null infinity is known as  a BMS$_3$ symmetry which is isomorphic to \eqref{algebra}. Thus we can find an interpretation for the charges $Q_{M_n}$ and $Q_{L_n}$ on the bulk side as the charges corresponding to the asymptotic symmetry generators. To be precise, let us consider a set of asymptotically flat spacetimes which transforms back into itself under the action of  asymptotic symmetry generators. In a particular coordinate systems,  known as BMS coordinates, the generic form of the asymptotically flat spacestimes with BMS$_3$ asymptotic symmetry is given by \cite{aspects}
 \begin{equation}\label{metric}
 ds^2=M du^2-2 du dr +2N du d\phi+r^2 d\phi^2,
 \end{equation}
 where 
 \begin{equation}
 M=\theta(\phi),\qquad N=\chi(\phi)+\dfrac{u}{2}\theta'(\phi),
 \end{equation}
 and $\theta(\phi)$ and $\chi(\phi)$ are arbitrary functions of the $\phi$ coordinate. $u$ is known as the \textit{retarded time} where for the Minkowski spacetime $u=t-r$. The generators of an infinitesimal  coordinate transformation, $\xi^\mu$, which preserve the form of the metric \eqref{metric},  are given by 
 \begin{equation}\label{inf.coord.trans}
 \xi^u=F,\qquad \xi^\phi=Y-{1\over r}\partial_\phi F,\qquad \xi^r=-r\partial_\phi Y+\partial_\phi^2 F-\frac{1}{r}N\partial_\phi F,
 \end{equation}
 where 
 \begin{equation}
 Y=Y(\phi),\qquad F=T(\phi)+u Y'(\phi),
 \end{equation}
 $Y(\phi)$ and $T(\phi)$ are arbitrary functions. $L_n$ and $M_n$ which are defined by 
 \begin{equation}\label{def.of. M.L}
 L_n=\xi(Y=ie^{in\phi},T=0),\qquad M_n=\xi(Y=0,T=ie^{in\phi}),
 \end{equation}
 satisfy the algebra \eqref{algebra} at large $r$. The corresponding charges of $L_n$ and $M_n$ can be computed by  various methods. They are given by  \textit{covariant phase space} method \cite{Barnich:2001jy},\cite{aspects} as\footnote{ The calculation  of surface charges in \cite{aspects} has been done at the circle at infinity. Moreover,  it is assumed that the background line element which is used to raise and lower  indices is Minkowski, $ds^2=-du^2-2 du dr +r^2 d\phi^2$.  } 
 \begin{align}\label{charges}
 \nonumber Q_{M_n}&={i\over 16\pi G} \int_0^{2\pi} d\phi \,e^{in\phi}\theta(\phi)+{i\over 8 G} \delta_n^0,\\
 Q_{L_n}&={i\over 8\pi G} \int_0^{2\pi} d\phi \,e^{in\phi}\chi(\phi).
 \end{align}
 The shift in the first line  of \eqref{charges} is necessary in order for the Poisson bracket of the charges produce the correct coefficient for the central term in the algebra \eqref{algebra}. The interesting point here  is that with this shift of charges we have $Q_{M_0}=Q_{L_0}=0$ for the Minkowski metric.  
 
  Substituting  \eqref{charges} in \eqref{stress tensor component field theory} one can find the components of the stress tensor as follows:
 \begin{align}\label{holographic T}
 \nonumber T^{uu}&=-{1\over 16 \pi G R }\left(1+\theta(\phi)\right),\\
 T^{u\phi}&={1\over 8 \pi G R^3 }\left(\chi(\phi)+\frac{u}{2}\theta'(\phi)\right).
 \end{align}
 This result is consistent with those of  \cite{Fareghbal:2013ifa} where the components of the stress tensor are calculated through   taking flat-space limit from the quasi-local stress tensor of the asymptotically AdS spacetimes. Moreover, we find the same results as in \cite{Bagchi:2015wna} if ${M}$ and ${N}$ in \cite{Bagchi:2015wna} are identified as the $T_{uu}$ and $T_{u\phi}$ components of the stress tensor. We have not fixed the constant  $R$  in the above calculations,yet. This can be done through relating  the constant term in the $uu$ component of the stress tensor with the  central charges of \eqref{algebra}.
 
By assuming a standard conservation formula for the components of the stress tensor one arrives at
 \begin{equation}
 \partial_u T^{u\phi}+\partial_\phi T^{\phi\phi}=0.
 \end{equation}
 Thus using \eqref{holographic T} we can determine $T^{\phi\phi}$  to be
 \begin{equation}
 T^{\phi\phi}=-{\theta(\phi)\over 16\pi G R^3 }+K,
 \end{equation}
 where $K$ is a constant of integration. If we also impose a traceless-ness condition $T^\mu_\mu=0$ for the stress tensor, $K$ is determined and we have
 \begin{equation}
  T^{\phi\phi}=-{1\over 16\pi G R^3 }\left(1+\theta(\phi)\right).
  \end{equation}
   
 From \eqref{holographic T} it is clear that the conservation equation, 
\begin{equation}
 \partial_u T^{uu}+\partial_\phi T^{\phi u}=0,
 \end{equation}
 is not satisfied for a symmetric stress tensor, i.e. $T^{u\phi}=T^{\phi u}$. One possible way to overcome  this obstacle is assuming a new conservation equation as $\partial_u T^{uu}=0$ \cite{Fareghbal:2013ifa}. However, if we want to write the conservation formula in a covariant way, there is a possibility of assuming non-symmetric stress tensors for the CCFTs. If we implement a non-symmetric stress tensor ( similar to the case in  \cite{Hartong:2015usd} ) such that $T^{u\phi}$ is non-zero and  is given by  \eqref{holographic T} but $T^{\phi u}=0$  then the holographic calculations result in the standard conservation equation, $\nabla_\mu T^{\mu\nu}=0$
 for the CCFT. The fact that CCFTs are not Poincare' invariant theories makes this assumption reliable. We should note again that all of these results are consequences of accepting a holographic duality between CCFTs and asymptotically flat spacetimes. In summary, we have
 \begin{align}\label{stress tensor final}
 \nonumber T_{uu}&=-{1\over 16 \pi G R }\left(1+\theta(\phi)\right),\\
\nonumber T_{u\phi}&=-{1\over 8 \pi G R  }\left(\chi(\phi)+\frac{u}{2}\theta'(\phi)\right),\\
\nonumber T_{\phi\phi}&=-{R\over 16\pi G  }\left(1+\theta(\phi)\right),\\
 T_{\phi u}&=0.
 \end{align}

 \section{Correlators of stress tensor}
 In this section we  use the results of the previous sections to calculate the correlation functions of CCFT$_2$. To do so, we assume that these functions are invariant   under the global part of the two\,- dimensional  symmetry algebra. For the two\,- dimensional theory, whose symmetry is given by \eqref{algebra}, the global part is generated by $\{L_0,L_{\pm 1}, M_0, M_{\pm 1}\}$. According to \eqref{stress tensor final}, the holographic calculations yield the components of stress tensor in terms of two functions $\theta(\phi)$ and $\chi(\phi)$. When we fix these functions on the gravity side, the asymptotically flat solution is completely determined. An infinitesimal coordinate transformation generated by \eqref{inf.coord.trans} changes these functions to $\theta+\delta\theta$  and $\chi+\delta\chi$. The infinitesimal changes of the functions can be calculated by using the Lie derivative of the metric components and expressing them in such a way that the generic form \eqref{metric} is preserved. We arrive at \cite{}
 \begin{align}\label{chnage.functons}
 \nonumber \delta_\xi\theta &=Y\theta'+2Y'\theta-2Y''',\\
 \delta_\xi \chi&=\frac{1}{2}T\theta'+Y\chi'+2Y'\chi+T'\theta-T'''.
 \end{align}
 We apply  \eqref{chnage.functons} on the gravity side to find the variation of the stress tensor in the boundary. Using \eqref{stress tensor final} and \eqref{chnage.functons} and imposing the conditions 
 \begin{equation}
 \delta_{M_n}\langle T_{ij}\rangle=0,\qquad\delta_{L_n}\langle T_{ij}\rangle=0,\qquad n=0,\pm1
 \end{equation}
 result in 
 \begin{equation}\label{1point}
 \langle T_{ij}\rangle=0,
 \end{equation}
 as expected.
 
  We can also   use \eqref{stress tensor final} and \eqref{chnage.functons} to calculate  higher-point functions. Since according to \eqref{stress tensor final},  $T_{\phi\phi}$ is the same as $T_{uu}$ up to an overall factor, its correlation functions with the other components are similar to the correlation functions of $T_{uu}$. Similar to the one-point functions, we want to determine the\textit{ p} point functions by imposing 
  \begin{equation}\label{condition2}
 \delta_{M_n}\langle T_{ij}^1\cdots T_{kl}^p\rangle=0,\qquad\delta_{L_n}\langle T_{ij}^1\cdots T_{kl}^p\rangle=0,\qquad n=0,\pm1
 \end{equation}
 where $T_{ij}^l=T_{ij}(u_l,\phi_l)$. 
 
 If we define  $\beta(\phi)=\theta(\phi)+1$ then the $uu$ and $\phi\phi$ components of the stress tensor will be proportional to $\beta(\phi)$. For     $ n=0,\pm\,1\ $ , equations \eqref{def.of. M.L} and \eqref{chnage.functons} yield the   following variations: 
 \begin{align}\label{var1}
\nonumber \delta_{M_n}\beta=0&,\qquad \delta_{L_n}\beta=e^{i n\phi}\left(i\partial_\phi\beta-2n\beta\right),\\
\delta_{M_n}\chi=\frac{1}{2}e^{i n\phi}\left(i\partial_\phi\beta-2n\beta\right)&,\qquad \delta_{L_n}\chi=e^{i n\phi}\left(i\partial_\phi\chi-2n\chi\right).
 \end{align}
 It is clear from \eqref{var1} that imposing $\delta_{L_n}\langle T_{ij}^1\cdots T_{kl}^p\rangle=0$ for $n=0,\pm\,1$  results in the equations 
 \begin{equation}\label{generic.equation}
 \sum_{k=1}^P\langle X_1\cdots e^{in\phi_k}(i\partial_k-2n)X_k\cdots X_p\rangle=0,
 \end{equation}
 where $X_i$ can be either  $\beta_i=\beta(\phi_i)$ or $\chi_i=\chi(\phi_i)$ and $\partial_k$ indicates the derivative with respect to the  $\phi$ at the point $\phi_k$. Thus we conclude that,  for a given $p$, all of the \textit{p} point functions of $\beta$ and $\chi$ with any numbers of $\beta$ and $\chi$ and any insertion of them have the same functionality  of $\{\phi_1,\phi_2,\cdots,\phi_p\}$ but with different overall constant factors. These constants can also be zero,  which  would render some correlation functions to vanish.

 The solution to Eq.  \eqref{generic.equation} is given by 
 \begin{equation}\label{generic solution}
 \langle X_1\cdots X_p\rangle=C{ e^{2i{\sum_{k=1}^p} \phi_k}\over \prod _{1\leq l<m\leq p}\left(e^{i\phi_l}-e^{i\phi_m}\right)^{4\over p-1}},
 \end{equation}
 where $C$ is a constant which can be zero. We determine $C$ by imposing $\delta_{M_n}\langle T_{ij}^1\cdots T_{kl}^p\rangle=0$ for $n=0,\pm\,1$. $\langle \chi_1\cdots \chi_p\rangle$ does not appear in any other equation; thus we conclude that it is given by the generic equation \eqref{generic solution}. Moreover, from $\delta_{M_n}\langle T_{u\phi}^1\cdots T_{u\phi}^p\rangle=0$ we find that 
 \begin{equation}
 \langle \beta_1\chi_2\chi_3\cdots \chi_p\rangle=\langle \chi_1\beta_2\chi_3\cdots \chi_p\rangle=\cdots=\langle \chi_1\chi_2\chi_3\cdots \beta_p\rangle.
 \end{equation}
 However, applying $\delta_{M_n}$ to other correlation functions which have at least one $T_{uu}$  reveals that the correlations of $\beta$ and $\chi$ containing more than one $\beta$ must be zero in order to be consistent with the generic result \eqref{generic solution}. Thus the only non-zero correlation functions of $\beta$ and $\chi$ are
 \begin{align}\label{final.beta.chi}
\nonumber \langle \chi_1\cdots \chi_p\rangle&=C_1{ e^{2i{\sum_{k=1}^p} \phi_k}\over \prod _{1\leq l<m\leq p}\left(e^{i\phi_l}-e^{i\phi_m}\right)^{4\over p-1}},\\ 
\langle \beta_1\chi_2\chi_3\cdots \chi_p\rangle=\cdots=\langle \chi_1\chi_2\chi_3\cdots \beta_p\rangle&=C_2{ e^{2i{\sum_{k=1}^p} \phi_k}\over \prod _{1\leq l<m\leq p}\left(e^{i\phi_l}-e^{i\phi_m}\right)^{4\over p-1}}
 \end{align}
 where $C_1$ and $C_2$ are constants. They must be related to the central charges $c_M$ and $c_L$ of \eqref{algebra} but on the gravity side they are only constants of integration. Now using \eqref{stress tensor final} and \eqref{final.beta.chi} we can calculate all of the correlation functions of the stress tensor:
 \begin{align}\label{correlation.final.generic}
\nonumber  \langle T_{uu}^1 T_{u\phi}^2\cdots T_{u\phi}^p\rangle &\propto C_2{ e^{2i{\sum_{k=1}^p} \phi_k}\over \prod _{1\leq l<m\leq p}\left(e^{i\phi_l}-e^{i\phi_m}\right)^{4\over p-1}}\\ 
  \langle T_{u\phi}^1 T_{u\phi}^2\cdots T_{u\phi}^p\rangle &\propto\left(C_1+{C_2\over 2}\sum_{k=1}^p u_k\partial_k\right) { e^{2i{\sum_{k=1}^p} \phi_k}\over \prod _{1\leq l<m\leq p}\left(e^{i\phi_l}-e^{i\phi_m}\right)^{4\over p-1}}
 \end{align}
  It is clear from  \eqref{correlation.final.generic} that all of  the\textit{ p} point functions of $T_{u\phi}$s are given by correlation of one $T_{uu}$ and $p-1$ of the $T_{u\phi}$.
 
 For the two-point functions we find 
\begin{align}\label{two-point-functions}
\nonumber \langle T_{uu}(\phi_1)T_{uu}(\phi_2)\rangle&=0,\\
\nonumber\langle T_{uu}(\phi_1)T_{u\phi}(u_2,\phi_2)\rangle&\propto C_2{e^{2i(\phi_1+\phi_2)}\over (e^{i\phi_1}-e^{i\phi_2})^4},\\
\langle T_{u\phi}(u_1,\phi_1)T_{u\phi}(u_2,\phi_2)\rangle&\propto C_1{e^{2i(\phi_1+\phi_2)}\over (e^{i\phi_1}-e^{i\phi_2})^4}\left(1+2i{C_2\over C_1}(u_2-u_1){e^{i\phi_1}+e^{i\phi_2}\over e^{i\phi_1}-e^{i\phi_2}}\right).
 \end{align}
which  are exactly the same as   results of \cite{Bagchi:2015wna} up to the constants $C_1$ and $C_2$ which are proportional to the central charges $c_M$ and $c_L$. Our results for the three-point functions are also exactly the same as \cite{Bagchi:2015wna}. However,  for the correlators  higher than four we cannot  regenerate the full correlation functions just by symmetry consideration similar to what we have done in this paper. In fact,  some cross-ratios  are necessary which we miss in this method. The main motivation for the calculations of this paper is to connect gravity calculations, using metric formulation in the three\,- dimensional asymptotically flat space times,  to the direct calculations of correlators in  Ref \cite{Bagchi:2015wna}. The consistency of our correlation functions (at least up to the four-dimensional correlators) with the results of \cite{Bagchi:2015wna} shows that a CCFT$_2$ could be a good candidate for the holographic dual of three\,- dimensional asymptotically flat spacetimes.
\section{Discussion }
In this paper we calculate all  correlation functions of CCFT$_2$ stress tensor by using Flat/CCFT proposal. On the gravity side we make use of the metric formulation of three-dimensional gravity.  Our method is applicable for higher-dimensional cases, as well. The interesting point is that the asymptotic symmetry of four-dimensional asymptotically flat spacetimes is also infinite-dimensional. The symmetry algebra is known as BMS$_4$ algebra. Using flat/CCFT correspondence we conclude that CCFT$_3$ also has an infinite\,- dimensional symmetry and therefore one expects universal behavior  for the correlation functions of the operators \cite{Bagchi:2016bcd}. Since the conformal symmetry in three dimensions is finite-dimensional, it is not clear how to find the correlators by taking the flat space limit. However, it is possible to generalize our  method in this paper  to the four-dimensional case. In the method which we use in this paper, the CCFTs are defined  by using their symmetries. These symmetries are given by BMS algebra which is infinite-dimensional in  two and three-dimensional CCFTs. In other words CCFTs are defined as  BMS-invariant field theories. Using this definition we can forget the contraction. The steps for the calculation of the CCFT$_3$ stress tensor correlators will be similar to those in two-dimensional case: The first step is to find the  stress tensor components by using the standard definition of conserved charges. The starting point is to  generalize   \eqref{2dcharge} to  three dimensions,  which can be performed very simply. However,  the main task is to derive the components by using the method of Sec.$ 2 $. We expect some sort of non-symmetric stress tensor components for the three-dimensional case. We emphasize again that in the derivation of stress tensor components we just need the direct connection  of CCFTs with the dual asymptotically flat spacetimes and parent CFT calculation is not necessary. The next step is to employ  the invariance of the correlators under the action of the global part of the BMS$_4$ symmetry ( as the symmetry of CCFT$_3$) which is expected to fix the structure of correlators.  Similar to CFT$_2$ and CCFT$_2$, it is plausible that the infinite-dimensional symmetry of CCFT$_3$ dictates some universal form for the \textit{n} point function of stress tensor correlators. If this is  so some related interesting  questions arise. Among them the issue of the entanglemet entropy of CCFT$_3$ which is expected to use the universality of the correlation  functions,  is of great interest\cite{Bagchi:2014iea}-\cite{Basu:2015evh}. 

Another interesting question in the context of Flat$_3$/CCFT$_2$ proposal is the calculation of the higher correlation functions by taking the appropriate  limit of the AdS/CFT calculations. This might be done along the lines  first  introduced in \cite{Costa:2013vza}. The importance of this problem is that its solution is a necessary step in the path   to find  holographic renormalization method  for the flat/CCFT correspondence. 
\subsubsection*{Acknowledgements}
The authors would like to thank A. Bagchi for useful comments. R. F would like to thank the School of Particles and Accelerator of Institute for Research in Fundamental Sciences (IPM) for the research facilities.

\appendix


\end{document}